\documentclass[12pt]{iopart}
\usepackage[english]{babel}
\usepackage[T1]{fontenc}
\usepackage[colorlinks=true,linkcolor=blue,citecolor=blue,urlcolor=blue]{hyperref}
\expandafter\let\csname equation*\endcsname\relax
\expandafter\let\csname endequation*\endcsname\relax
\usepackage{bm,bbm,amssymb,amsmath}
\usepackage{graphicx}
\usepackage{multirow,booktabs,dcolumn}
\usepackage{slashed}
\usepackage{isotope}
\usepackage[capitalize]{cleveref}
\usepackage{nicefrac}
\usepackage{ulem}

\usepackage[sort&compress,numbers,square]{natbib}

\begin{document}

\title[Exact restoration of Galilei invariance in DFT with QMC]{Exact restoration of Galilei invariance in density functional calculations with quantum Monte Carlo}

\author{P. Massella}
\address{Physics Department, University of Trento, via Sommarive 14, I-38123 Trento, Italy}

\author{F. Barranco}
\address{Departamento de Fisica Aplicada III, Escuela Superior de Ingenieros,
Universidad de Sevilla, Camino de los Descubrimientos, Sevilla, Spain}

\author{D. Lonardoni}
\address{Facility for Rare Isotope Beams, Michigan State University, East Lansing, Michigan 48824, USA}
\address{Theoretical Division, Los Alamos National Laboratory, Los Alamos, New Mexico 87545, USA}

\author{A. Lovato}
\address{INFN-TIFPA Trento Institute of Fundamental Physics and Applications, 38123 Trento, Italy}
\address{Physics Division, Argonne National Laboratory, Argonne, IL 60439}

\author{F. Pederiva}
\address{Physics Department, University of Trento, via Sommarive 14, I-38123 Trento, Italy}
\address{INFN-TIFPA Trento Institute of Fundamental Physics and Applications, 38123 Trento, Italy}

\author{E. Vigezzi}
\address{INFN Sezione di Milano, Via Celoria 16, I-20133 Milano, Italy}
\ead{enrico.vigezzi@mi.infn.it}

\vspace{10pt}
\begin{indented}
\item[]June 2019
\end{indented}

\begin{abstract}
Galilean invariance is usually violated in self-consistent mean-field calculations that employ effective density-dependent nuclear forces. We present a novel approach, based on variational quantum Monte Carlo techniques, suitable to preserve this symmetry and assess the effect of its violation, seldom attempted in the past. To this aim, we generalize the linear optimization method to encompass the density-dependence of effective Hamiltonians, and study \isotope[4]{He}, \isotope[16]{O}, and \isotope[40]{Ca} ground-state properties employing the Gogny interaction.
\end{abstract}

\section{Introduction}
\label{sec:intro}

Respecting Galilean invariance in nuclear structure calculations represents a challenging problem, mostly due to the need to antisymmetrize the wave functions in order to respect the Pauli principle. In the case of light systems, various techniques have been adopted to treat this problem exactly, like the Faddeev~\cite{Witala:2001}, Faddeev-Yakubovsky~\cite{Lazauskas:2004,Lazauskas:2009}, Alt-Grassberger and Sandhas~\cite{Deltuva:2007}, and hyperspherical harmonics~\cite{Kievsky:2008,Marcucci:2009} methods, as well as the no-core shell model approach~\cite{Navratil:2000,Barrett:2013}, recently extended to deal with the continuum and with scattering problems~\cite{Baroni:2013,Navratil:2016}.

On the other hand, in most calculations for heavier nuclei the antisymmetrization is carried out by expanding the wave functions in terms of Slater determinants, which depend on all the $3A$ coordinate of the system and therefore induce a contamination due to the motion of the system as a whole. This effect is mitigated by subtracting the kinetic energy of the center of mass (CM) from the Hamiltonian, and exploiting the fact that symmetry breaking effects are expected to scale as $1/A$. Furthermore, approaches based on self-consistent mean-field theory generally make use of effective interactions which contain density-dependent terms which explicitly violate Galilei invariance~\cite{Ring:1980,Bender:2003}.

An extensive and very interesting analysis of this issue was carried out in a series of papers by K.W.~Schmid {\it et al}~\cite{Schmid:2001,Schmid:2002a,Schmid:2002b,Schmid:2003,Rodriguez:2004a,Rodriguez:2004b}, who were able to perform exact variation after projection (VAP) calculations for heavy nuclei, typically studied by self-consistent mean-field theory. Schmid obtained analytic results using oscillator wave functions with the Gogny interaction~\cite{Decharge:1979}, modifying the functional dependence of its density-dependent term, so that it does not explicitly violate translation invariance~\cite{Schmid:2002b}. The complexity of the approach prevented however the use of the Gogny interaction in the case of Hartree-Fock wave functions, and results were only obtained using the schematic, density-independent Brink-Boeker interaction~\cite{Rodriguez:2004a,Rodriguez:2004b}. These studies showed that fulfilling Galilean invariance leads to non-trivial consequences concerning not only nuclear binding energies, but also the properties of form factors and spectral functions. In spite of their general interest, these calculations remain up to date the only available quantitative studies of the problem, see \cite{Robledo:2018cdj} for a recent review.

In this work, we present a different approach to the issue, based on numerical quantum Monte Carlo (QMC) techniques. One of the great advantages of QMC is that one can directly access the $3A$ coordinates of the nucleons, so that it is easy to swap between a fixed reference frame and the CM frame. The combination of algorithmic developments and the increasing availability of computational resources, makes it possible to perform reliable QMC calculations for nuclei of mass up to $A \approx 50$ with density-independent Hamiltonians. On the other hand, solving a self-consistent equation, defined by the use of a density-dependent term, represents a numerical challenge within current QMC algorithms. In a variational Monte Carlo (VMC) calculation~\cite{Carlson:2015} the wave function is parametrized assuming some given analytic form. The development of automatic minimization methods, such as the Linear Method (LM) introduced by Toulouse and Umrigar~\cite{Toulouse:2007}, allows for the optimization of wave functions characterized by a large number of variational parameters, which translates into the possibility of reaching very accurate solutions~\cite{Contessi:2017}. 

Building on such progress, it is now possible to apply the QMC formalism to address self-consistent density functional calculations. By generalizing the LM to treat density-dependent interactions, we are in the position to bring the analysis by Schmid {\it et al} a step further. We solve the density functional problem self-consistently employing accurate effective interactions, like Gogny D1S~\cite{Berger:1991}, fully respecting Galilean invariance and carefully elucidating the effects of its violation. The QMC approach makes it possible to project the Hartree-Fock (HF) wave functions in the CM frame (projection after variation, PAV), or to directly optimize the wave functions in the CM frame (VAP). 

The paper is organized as follows: in section~\ref{sec:gogny} the Gogny force is reviewed in the context of mean-field calculations, later discussed in section~\ref{sec:mf}; section~\ref{sec:qmc} deals with quantum Monte Carlo methods, in particular with the variational Monte Carlo algorithm and the linear method for wave function optimization; results for the comparison and analysis of VMC vs. HF calculations are presented in section~\ref{sec:res}; section~\ref{sec:con} is devoted to conclusions and to possible extensions of the present work.

\section{Gogny interaction}
\label{sec:gogny}
In this work, calculations are performed using the Gogny effective interaction~\cite{Decharge:1979}. The parameters of the interaction have been fit to a few experimental properties, comparing empirical data in spherical nuclei with calculations performed within the Hartree-Fock approximation (see below). We will only consider the \isotope[4]{He}, \isotope[16]{O}, and \isotope[40]{Ca} nuclei, for which no pairing interaction needs to be included
in the calculation. The associated non-relativistic Hamiltonian is the sum of the kinetic term and a two-body, density-dependent interaction:
\begin{align}
	H=-\sum_{i=1}^{A} \frac{\nabla_i^2}{2m_N}+\sum_{i<j}^{A}v_{ij},
\label{eq:Ham}
\end{align}
where $m_N$ is the nucleon mass, and derivatives are calculated with respect to the nucleon coordinates.

Neglecting spin-orbit contributions, which are not relevant for the aims of the present work, the two-body potential is defined as a sum of three terms
\begin{align}
v_{ij}=v^4_{ij}+v^{\rm ddp}_{ij}+v^{\rm C}_{ij}.
\label{eq:gogny}
\end{align}
The first contribution has a spin-isospin structure analogous to that of the first four components of the Argonne $v_{18}$ interaction~\cite{Wiringa:1995} 
\begin{align}
v^4_{ij}=\sum_{k=1,2}(W_k+B_k\,\mathcal P_\sigma-H_k\,\mathcal P_\tau-M_k\,\mathcal P_\sigma\,\mathcal P_\tau)\,e^{-r_{ij}^2/\mu_k^2},
\label{eq:gogny_1}
\end{align}
where the spin/isospin-exchange operators take the form
\begin{align}
\mathcal P_\sigma=\dfrac{1+\sigma_{ij}}{2}, \qquad \mathcal P_\tau=\dfrac{1+\tau_{ij}}{2} ,
\end{align}
and $\sigma_{ij}=\bm\sigma_i\cdot\bm\sigma_j$ and $\tau_{ij}=\bm\tau_i\cdot\bm\tau_j$ are the scalar products between the spin and isospin matrices of the $i$th and $j$th particles. The second contribution is a zero-range density-dependent term, given by
\begin{align}
v^{\rm ddp}_{ij}= \tau_0\,(1+x_0\,\mathcal P_\sigma)\,\rho^\alpha({\bf R}_{ij})\,\delta({\bf r}_{ij}).
\label{eq:gogny_2}
\end{align}
The third term $v^{\rm C}_{ij}$ denotes the Coulomb potential. In the above equations ${\bf r}_{ij}={\bf r}_i-{\bf r}_j$ and ${\bf R}_{ij}=({\bf r}_i+{\bf r}_j)/2$ are the relative and CM coordinate of the nucleon pair. We shall make use of the D1S parametrization of the Gogny interaction introduced in \cite{Berger:1991} for the calculation of fission processes and extensively used in the literature~\cite{Goutte:2005da,RocaMaza:2011pm,Gaffney:2013,Sellahewa:2014}. The parameters defining the density-independent part of the interaction are listed in table~\ref{tab:gogny}, while the values of the coefficients entering the density-dependent component are $\alpha=1/3$, $\tau_0=1390.6\,\rm MeV$, and $x_0=1$. 

\begin{table}[b]
\centering
\caption[]{Parameters of the density-independent contributions of the Gogny D1S interaction as from \cite{Berger:1991}. All quantities are in MeV, except for $\mu_k$ that is in fm.}
\begin{tabular}{cccccc}
$k$ & $W_i$    & $B_k$    & $H_k$    & $M_k$    & $\mu_k$ \\
\hline
$1$ & $-1720.30$ & $1300.0$ & $-1813.53$ & $1397.60$ & $0.7$ \\
$2$ & $\phantom{-1}103.64$ & $-163.48$ & $\phantom{-1}162.81$  & $-223.93$ & $1.2$ \\
\end{tabular}
\label{tab:gogny}
\end{table}

Note that, in order to be implemented in a QMC algorithm, the delta function entering $v^{\rm ddp}_{ij}$ is smeared introducing a Gaussian regulator
\begin{align}
\delta(r_{ij})\to G(r_{ij})=\dfrac{1}{(\mu_3\sqrt{\pi})^3}\,e^{-r_{ij}^2/\mu_3^2} .
\label{eq:ddp_reg}
\end{align}
We have carefully analyzed the cutoff dependence of our results, by varying the Gaussian regulator between $\mu_3=0.4\,\rm fm$ and $\mu_3=0.1\,\rm fm$. Since finite-size effects are found to be small for $\mu_3\lesssim0.15\,\rm fm$, we performed our quantum Monte Carlo calculations keeping $\mu_3=0.1\,\rm fm$. The regulator used in the present work is significantly harder than those typically adopted to regularize the contact terms entering local chiral forces~\cite{Piarulli:2014,Piarulli:2017}, which lie in the range $0.6 - 0.8\,\rm fm$. Choosing such a hard regulator, while leading to additional difficulties in solving the many-body problem---though still manageable with QMC methods---has the advantage of preventing the appearance of the so-called {\it regulator artifacts} (see \cite{Lovato:2011ij,Huth:2017} and references therein for details). Note that a similar regularization procedure should be followed to include the spin-orbit component of the Gogny interaction in our QMC method.

\section{Mean-field theory}
\label{sec:mf}

\subsection{Hartree-Fock}
\label{sec:hf}
Within the HF approach, the nuclear wave function is assumed to be a Slater determinant  $|\Phi\rangle$ formed by a set of single-particle wave functions $\chi_{\alpha}$, where $\alpha$ stands for the spherical quantum numbers:
\begin{align} 
\langle X|\Phi\rangle = \mathcal A\left\{\prod_{\alpha=1}^A \chi_\alpha(x_i)\right\} .
\label{eq:phi}
\end{align}
In the above equation, $\mathcal A$ is the antisymmetrization operator and $|X\rangle=|x_1,\ldots,x_A\rangle$, and $x_i=\{{\bf r}_i,\sigma_i,\tau_i\}$ are generalized coordinates representing position, spin, and isospin of the $i$th nucleon. 

The nuclear mean-field is determined by finding the Slater determinant $|\Phi\rangle$ that minimizes the energy 
\begin{equation}
E^{\rm HF}(\rho)=\langle\Phi|H|\Phi\rangle ,
\end{equation}
which is a functional of the single-particle density 
\begin{equation}
\rho_{\alpha\alpha'}= \langle \Phi | a^{\dagger}_{\alpha'} a_{\alpha} | \Phi \rangle, 
\end{equation}
written in terms of the creation and annihilation operators $a^{\dagger}_\alpha, a_\alpha$.

Following \cite{Ring:1980}, the solution of such a minimization defines the HF single-particle basis $\phi_{a}$ and the HF average potential
\begin{align}
H^{\rm HF} =\sum_i^A h(i),
\end{align}
where $h$ depends on the density and  obeys the relation
\begin{align}
h_{aa'} = \frac{\partial E^{\rm HF}(\rho)}{\partial \rho_{a'a}}=
\epsilon_a\,\delta_{aa'}.
\end{align}
The HF solution provides a single-particle basis in which both $h$ and $\rho$ are diagonal.

In order to determine the mean-field wave functions, they are expanded in a spherical Woods-Saxon basis $\chi_\alpha$
\begin{align}
\phi_a = \sum_{\alpha} D_{\alpha a}\,\chi_{\alpha},
\end{align}
so that
\begin{align}
\rho_{\alpha\alpha'}=\sum_{a=1}^A D_{\alpha a}D^*_{\alpha'a}    
\end{align}
The self-consistent HF equations can then be written as an eigenvalue problem (see \cite{Ring:1980}, Eq.~(5.38)) 
\begin{align}
\sum_{\alpha'} h_{\alpha\alpha'}\,D_{\alpha'a} = \epsilon_a\,D_{\alpha a},
\end{align}
where the matrix elements $h_{\alpha\alpha'}$ are given by
\begin{align}
h_{\alpha\alpha'} =& \;t_{\alpha\alpha'} + 
\sum_{\beta\beta'}\Bigg(\bar v_{\alpha\beta'\alpha'\beta}+\frac{1}{2}\sum_{\gamma\gamma'}\Big\langle\gamma'\beta'\Big|\frac{\partial \bar v}{\partial \rho_{\alpha'\alpha}}\Big|\gamma \beta\Big\rangle\,\rho_{\gamma\gamma'}\Bigg)  \rho_{\beta\beta'},
\label{eq:hf_iter}
\end{align}
and $t$ and $\bar v$ denote, respectively, the matrix elements of the kinetic energy and the antisymmetrized matrix elements of the two-body interaction. These equations must be solved by iteration, assuming an initial choice for the transformation coefficients $D_{\alpha}$. In the case of density-dependent interactions, the matrix element $\bar v$ must be recalculated at each iteration, and furthermore the rearrangement term $\partial \bar v /\partial \rho$ must be taken into account. The Woods-Saxon basis used to compute the matrix elements of the two-body interaction ${\bar v}$ is obtained by solving the Schr\"odinger equation for a Woods-Saxon potential of standard form~\cite{Bohr:1969} in a box of $20\,\rm fm$, using an energy cutoff of $400\,\rm MeV$~\cite{Briganti:1998}. The use of a Woods-Saxon basis is particularly convenient in the case of weakly-bound systems, but it is equivalent to the use of a harmonic oscillator basis for the case of the well bound nuclei considered here.  

\subsection{Effective interaction: fitting procedure}
\label{sec:fit}
According to the fitting procedure adopted to establish the parameters of the Gogny interaction~\cite{Decharge:1979}, HF calculations are performed correcting for the violation of translation invariance by subtracting the kinetic CM contribution $T_{\rm CM}$ from the Hamiltonian, where 
\begin{align}
T_{\rm CM}=\frac{P^2_{\rm CM}}{2M}=-\frac{1}{A} \sum_{i=1}^A\frac{\nabla^2_i}{2m_N}- \frac{1}{A} \sum_{i\ne j}^{A}\frac{\bm{\nabla}_i\cdot\bm{\nabla}_j}{2m_N},
\label{eq:ke_cm}
\end{align}
and the mass of the nucleus is given by $M=A\,m_N$. The first term, a one-body operator, can be dealt with by simply rescaling the total kinetic energy:
\begin{align}
-\sum_{i=1}^A\frac{\nabla_i^2}{2m_N}\rightarrow - \left(1-\frac{1}{A}\right)\sum_{i=1}^A\frac{\nabla_i^2}{2m_N}.
\end{align}
On the other hand, the last term of equation~(\ref{eq:ke_cm}) is a two-body operator, and it has to be treated on the same footing as the two-body components of the potential of equation~(\ref{eq:gogny}). We remark that in the fitting protocol of other effective interactions, only the one-body part of the kinetic energy correction is taken into account.

In spite of this subtraction procedure, the resulting wave functions still violate the basic translational symmetry. While symmetry breaking in nuclear physics has been the subject of an immense literature, in the case of translational invariance (see \cite{Engel:2007,Messud:2011,Messud:2013,Lesinski:2014} for recent studies) there are surprisingly few quantitative estimates of the errors due to the violation of this symmetry in HF calculations. This is mostly due to the fact that mean-field theory is mainly applied to medium-heavy nuclei, for which this violation, which is expected to scale as $1/A$, is small. A remarkable exception is represented by a series of papers by K.W.~Schmid, who calculated the binding energies and radii of few closed-shell nuclei, including \isotope[4]{He}, \isotope[16]{O}, and \isotope[40]{Ca}, fully restoring the Galilean invariance in the case of density-independent interactions by making use of the analytic properties of harmonic oscillator configurations~\cite{Schmid:2002b}. It should be noted that the density-dependent part of the Gogny interaction $v^{\rm ddp}_{ij}$, that depends on the density at the center of mass of the nucleon pair ${\bf R}_{ij}$, explicitly violates Galilean invariance. Schmid modified this dependence using the density $\rho^\alpha({\bf R}_{ij}-{\bf R}_{\rm CM})$ instead of $\rho^\alpha({\bf R}_{ij})$ (see also the discussion about internal density in \cite{Giraud:2008}). This procedure restores Galilean invariance, but it introduces a $A$-body force that makes HF calculations much more difficult to perform. On the other hand, this modification is suitable for QMC calculations, as discussed in the next section.

\section{Quantum Monte Carlo}
\label{sec:qmc}

\subsection{Variational Monte Carlo}
\label{sec:vmc}
In the VMC method~\cite{Carlson:2015}, provided a trial wave function $\Psi_T$, the expectation value of the Hamiltonian is given by
\begin{align}
E_V=\langle H\rangle=\frac{\langle\Psi_T|H|\Psi_T\rangle}{\langle\Psi_T|\Psi_T\rangle}\geq E_0 ,
\label{eq:ev}
\end{align}
where $E_0$ is the energy of the true ground state with the same quantum numbers as $\Psi_T$, and the rightmost equality is valid only if the wave function is the exact ground-state wave function $\Psi_0$. The energy expectation value of equation~(\ref{eq:ev}) typically depends on the quality of the employed wave function. In the VMC method, one minimizes $E_V$ with respect to changes in the variational parameters, in order to obtain $\Psi_T$ as close as possible to $\Psi_0$.

In this work we employ a trial wave function of the form
\begin{align}
\Psi_T(X)\equiv\langle X|\Psi_T\rangle=\langle X|\Phi\rangle,
\label{eq:psi}
\end{align}
where $\langle X|\Phi\rangle$ is the same Slater determinant of equation~(\ref{eq:phi}). In our VMC calculations, the single-particle states are taken to be
\begin{align} 
\phi_a(x_i) = R_{nl}(r_i)\,Y_{ll_z}(\hat r_i)\,Y_{ss_z}(\sigma)\,Y_{tt_z}(\tau) ,
\label{eq:spo}
\end{align}
where $R_{nl}(r)$ is the radial function, $Y_{ll_z}$ is the spherical harmonic, and $Y_{ss_z}(\sigma)$ and $Y_{tt_z}(\tau)$ are the complex spinors describing the spin and isospin of the single-particle state.

As described in \cite{Contessi:2017}, the radial functions $R_{nl}(r)$, are expressed as a sum of cubic splines, characterized by a smooth first derivative and a continuous second derivative. The large number of variational parameters involved in the construction of such radial components allows enough flexibility to obtain very accurate trial wave functions, with optimal variational energies very close to those calculated by performing the imaginary-time propagation (see \cite{Carlson:2015} and references therein for more details).

Note that, using the wave function of equation~(\ref{eq:psi}) and inserting a completeness over the generalized coordinate $X$, the expectation value of a generic operator $\mathcal O$ can be expressed as
\begin{align}
\langle \mathcal O \rangle & = \frac{\langle\Psi_T|\mathcal O|\Psi_T\rangle}{\langle\Psi_T|\Psi_T\rangle} =
\dfrac{\sum_X P(X)\, O_L(X)}{\sum_X P(X)} ,
%\dfrac{\sum_X \left|\Psi_T(X)\right|^2 \dfrac{\langle X |\mathcal O |\Psi_T\rangle}{\langle %X|\Psi_T\rangle}}{\sum_X \left|\Psi_T(X)\right|^2} ,
\label{eq:ave}
\end{align}
where $P(X)=|\Psi_T(X)|^2$ can be interpreted as a probability distribution of points $\{X\}$ in a multidimensional space, and 
\begin{align}
    O_L(X)=\dfrac{\langle X |\mathcal O |\Psi_T\rangle}{\langle X|\Psi_T\rangle} ,
    \label{eq:local}
\end{align}
is referred to as the local expectation value. Equation~(\ref{eq:ave}) actually corresponds to a multidimensional integral that can be calculated using Monte Carlo sampling. According to the Metropolis algorithm~\cite{Metropolis:1953}, a number of configurations $X_i$ are sampled from the probability distribution $P(X)$, and the expectation value of the operator $\mathcal O$ is calculated as
\begin{align}
%\langle \mathcal O\rangle_L=\frac{1}{\mathcal M}\sum_{i=1}^{\mathcal M} \frac{\langle X_i|\mathcal O|\Psi_T\rangle}{\langle X_i|\Psi_T\rangle} ,
\langle \mathcal O\rangle=\frac{1}{\mathcal M}\sum_{i=1}^{\mathcal M} O_L(X_i) ,
\label{eq:MCave}
\end{align}
where $\mathcal M$ is the number of sampled configurations. Details on the sampling procedure and Monte Carlo statistical errors can be found, e.g. in \cite{Ceperley:1995}.

Note that, in order to remove CM contributions, the nuclear wave function and the observables are calculated in the CM rest frame (COMF in the following) instead than in a fixed rest frame (FRF in the following), subtracting the CM position from all the spatial coordinates:
\begin{align}
{\bf r}_i\to{\bf r}_i-{\bf R}_{\rm CM},\qquad{\bf R}_{\rm CM}=\frac{1}{A}\sum_{j=1}^A{\bf r}_j.
\label{eq:r_i}
\end{align}
By introducing the ``intrinsic'' coordinate ${\bf r}^{\rm int}_i={\bf r}_i-{\bf R}_{\rm CM }$, the above procedure is equivalent to replacing the radial function and the spherical harmonic of equation~(\ref{eq:spo}):
\begin{equation}
R_{nl}(r_i) \to R_{nl}(r_i^{\rm int}) ,\qquad Y_{ll_z}(\hat{{\bf r}}_i) \to Y_{ll_z}(\hat{{\bf r}}_i^{\rm int }).
\end{equation}
Each single-particle orbital depends now upon the coordinates of all the nucleons, in a way that resembles backflow correlations~\cite{Schmidt:1981}. Hence, although the trial wave function is constituted by a single Slater determinant, subtracting the CM coordinate brings about some degree of correlations among the nucleons, whose importance decreases with $A$. As a consequence, the determination of the radial wave functions that minimize the expectation value of the energy becomes a numerically challenging problem, because it involves the calculation of non-factorizable integrals in $3A$ dimensions. However, this problem lends itself to a solution based on QMC techniques, following the iterative procedure outlined in the next section.

\subsection{Linear optimization method}
\label{sec:lm}
In order to optimize the radial components of the trial wave function, in this work we adopt the {\it linear method} (LM)~\cite{Toulouse:2007}, that was applied for the first time in a nuclear quantum Monte Carlo calculation in \cite{Contessi:2017}. Let us first define the normalized trial variational state
\begin{align}
| \bar{\Psi}_T ({\bf p}) \rangle = \frac{| \Psi_T ({\bf p}) \rangle}
{\sqrt{\langle \Psi_T ({\bf p}) | \Psi_T ({\bf p}) \rangle }},
\end{align}
as a function of the $\mathcal N_p$ variational parameters ${\bf p}=\{p_1,\dots,p_{\mathcal N_p}\}$.

Within the LM, at each optimization step one performs a first-order expansion around the current set of variational parameters ${\bf p}^0$
\begin{align}
 | \bar{\Psi}_T^{\rm lin}({\bf p} ) \rangle = 
 | \bar{\Psi}_T^0 ({\bf p}^0) \rangle 
 + \sum_{i=1}^{\mathcal N_p} \Delta p_i \,|\bar{\Psi}_T^i ({\bf p}^0) \rangle ,
\label{eq:lm_expansion}
\end{align}
where $|\bar{\Psi}_T^0 ({\bf p}^0)\rangle\equiv|\Psi_T ({\bf p}^0)\rangle$, and for $i>0$ 
\begin{align}
 |\bar{\Psi}^i_T ({\bf p}^0) \rangle &
 =\frac{\partial | \bar{\Psi}_T({\bf p}) \rangle}{\partial p_i}
  \Big|_{{\bf p}={\bf p}^0}\nonumber\\
 &= |\Psi_T^i({\bf p}^0) \rangle - S_{0i}|\Psi_T({\bf p}^0) \rangle .
\end{align}
In the latter equation the first derivative with respect to the $i$th parameter is given by
\begin{align}
|\Psi_T^i ({\bf p}^0 ) \rangle = \frac{ \partial |\Psi_T({\bf p})\rangle}
{\partial p_i}\Big|_{{\bf p}={\bf p}^0},
\end{align}
while the overlap matrix is defined as 
\begin{align}
S_{0i}=\langle \Psi_T ({\bf p}^0) | \Psi_T^i  ({\bf p}^0)\rangle.
\end{align}
By using the normalization freedom we can impose $\langle \bar{\Psi}_T^0 ({\bf p}^0)|  \bar{\Psi}_T^0 ({\bf p}^0)\rangle=1$, so that the derivatives of $| \bar{\Psi}_T({\bf p}) \rangle$ are orthogonal to 
$| \bar{\Psi}_T^0({\bf p}^0) \rangle$
\begin{align}
\langle  \bar{\Psi}_T^0 ({\bf p}^0)|  \bar{\Psi}^i_T  ({\bf p}^0)  \rangle=0 .
\end{align}
The eigenvalue equation for the Hamiltonian in the basis formed by the $(\mathcal N_p+1)$-dimensional basis $\Big\{|\bar{\Psi}_T^0({\bf p}^0)\rangle,  
\dots, |\bar{\Psi}^{\mathcal N_p}_T({\bf p}^0)\rangle\Big\}$ reads
\begin{align}
H\, \sum_{j=0}^{\mathcal N_p} \Delta {\bf p}^j\,|\bar{\Psi}^j_T({\bf p}^0)\rangle  = 
E\, \sum_{j=0}^{\mathcal N_p} \Delta {\bf p}^j\,|\bar{\Psi}^j_T({\bf p}^0)\rangle .
\end{align}
Multiplying the latter equation by $\langle \bar{\Psi}^i_T({\bf p}^0)|$ yields to the generalized eigenvalue equation
\begin{align}
\sum_j \bar{H}_{ij} \, \Delta {\bf p}^j = E \sum_j \, \bar{S}_{ij} \, \Delta {\bf p}^j ,
\label{eq:lin_sys}
\end{align}
where the Hamiltonian and overlap matrix elements are
\begin{align}
\bar{H}_{ij} &= \langle \bar{\Psi}^i_T({\bf p}^0)| H | \bar{\Psi}^j_T({\bf p}^0)\rangle, \nonumber\\
\bar{S}_{ij} &= \langle \bar{\Psi}^i_T({\bf p}^0)| \bar{\Psi}^j_T({\bf p}^0)\rangle .
\end{align}

The linear method consists of solving equation~(\ref{eq:lin_sys}) for the lowest eigenvalue and associated eigenvector $\Delta\bar{{\bf p}}$. It has to be noted that equation~(\ref{eq:lin_sys}) can alternatively be obtained by minimizing the energy expectation value on the linear wave function
\begin{align}
E_{\rm lin}({\bf p})\equiv \frac{\langle\bar{\Psi}_T^\text{lin}({\bf p})|
H  |\bar{\Psi}_T^{\rm lin}({\bf p})\rangle}
{\langle\bar{\Psi}_T^{\rm lin}({\bf p})|\bar{\Psi}_T^{\rm lin}({\bf p})
\rangle} ,
\end{align}
with respect to changes in the variational parameters~\cite{Toulouse:2007}.

The expressions of the above matrix elements for real variational parameters can be found in \cite{Umrigar:2005}. Here we report their expressions for the complex case~\cite{Motta:2015}, needed in nuclear QMC calculations. Inserting a completeness over the generalized coordinate $X$, the Hamiltonian and overlap matrix elements read (for brevity the dependence on ${\bf p}^0$ is understood)
\begin{align}
\bar{H}_{ij}&=\sum_X \langle \bar{\Psi}^i_T| X\rangle \langle X | H | \bar{\Psi}^j_T \rangle ,\nonumber\\
\bar{S}_{ij}&=\sum_X \langle \bar{\Psi}^i_T| X\rangle \langle X | \bar{\Psi}^j_T \rangle .
\end{align}
By making explicit the definition of $ | \bar{\Psi}^i_T \rangle$ one obtains
\begin{align}
\langle X | H |\bar{\Psi}^i_T\rangle &= \langle X | H | \Psi^i_T\rangle - S_{0i}\langle X | H | \Psi_T \rangle , \nonumber \\
\langle X | \bar{\Psi}^i_T\rangle &= \langle X| \Psi^i_T\rangle - S_{0i} \langle  X | \Psi_T  \rangle ,
\end{align}
where the matrix element $\langle X | H |\Psi^i_T\rangle$ can be expressed in terms of the the local energy $E_L(X)=\langle X|H|\Psi_T\rangle/\langle X|\Psi_T \rangle$ (see equation~(\ref{eq:local})) and its derivative $E_L^i(X)$:
\begin{align}
\langle X | H |\Psi^i_T\rangle = \langle X | \Psi_T\rangle E_L^i(X) + \langle X | \Psi^i_T\rangle E_L(X).
\end{align}
The derivatives of the wave function and of the local energy are numerically evaluated using the three-point stencil rule by shifting the $i-$parameter by a small quantity $\delta_p$: $p_i \to p_i \pm \delta_p$. 

The evaluation of $S_{0i}$ requires inserting an additional completeness relation
\begin{align}
S_{0i} = \sum_X  \langle \Psi_T | X\rangle \langle X| \Psi^i_T\rangle
=\Big\langle \frac{\Psi^i_T}{\Psi_T} \Big\rangle ,
\end{align} 
where the rightmost term is intended as in equation~(\ref{eq:ave}). 

Collecting the above results, the Hamiltonian and overlap matrix elements can then be expressed as an average over finite Monte Carlo samples, as in equation~(\ref{eq:MCave}):
\begin{align}
\bar{H}_{00}&=\Big\langle E_L \Big\rangle, \nonumber\\
\bar{H}_{0i}&= \Big\langle E_L^i \Big\rangle +\Big\langle \frac{\Psi^i_T}{\Psi_T}E_L\Big\rangle -  \Big\langle \frac{\Psi^i_T}{\Psi_T}  \Big\rangle \Big\langle E_L \Big\rangle, \nonumber\\
\bar{H}_{i0}&=\Big\langle \frac{\Psi^{i*}_T}{\Psi_T}E_L\Big\rangle - \Big\langle \frac{\Psi^{i*}_T}{\Psi_T} \Big\rangle \Big\langle E_L \Big\rangle, \nonumber\\
\bar{H}_{ij}&=\Big\langle \frac{\Psi^{i*}_T}{\Psi_T}  E_L^j \Big\rangle + \Big\langle \frac{\Psi^{i*}_T}{\Psi_T} \frac{\Psi^j_T}{\Psi_T} E_L \Big\rangle 
- \Big\langle \frac{\Psi^{i*}_T}{\Psi_T} E_L \Big\rangle \Big\langle \frac{\Psi^j_T}{\Psi_T} \Big\rangle \nonumber\\
&- \Big\langle E_L^j \Big\rangle \Big\langle \frac{\Psi^{i*}_T}{\Psi_T} \Big\rangle - \Big\langle  \frac{\Psi^j_T}{\Psi_T} E_L\Big\rangle
 \Big\langle \frac{\Psi^{i*}_T}{\Psi_T} \Big\rangle + \Big\langle E_L \Big\rangle \Big\langle  \frac{\Psi^j_T}{\Psi_T} \Big\rangle \Big\langle  \frac{\Psi^{i *}_T}{\Psi_T} \Big\rangle, \nonumber\\
\bar{S}_{00}&=1, \nonumber\\
\bar{S}_{0i}&=\bar{S}_{i0}=0, \nonumber\\
\bar{S}_{ij}&=\Big\langle \frac{\Psi^{i*}_T}{\Psi_T}  \frac{\Psi^j_T}{\Psi_T}  \Big\rangle -
\Big\langle \frac{\Psi^{i*}_T}{\Psi_T}\Big\rangle \Big\langle \frac{\Psi^j_T}{\Psi_T}\Big\rangle.
\label{eq:lm_est}
\end{align}

The hermiticity of the matrix $\bar{H}$ is recovered only in the limit of an infinite Monte Carlo sample. However, even for a finite sample, we refrain from symmetrizing $H_{ij}$, as employing non Hermitian estimators leads to a stronger zero-variance principle~\cite{Nightingale:2001}, and to smaller statistical errors on a finite sample than using its symmetrized analog~\cite{Umrigar:2005}. It has been shown that the Monte Carlo statistical uncertainty is largely reduced by the fact that the above matrix elements are written in terms of covariances~\cite{Umrigar:2005}. Nevertheless, for a finite sample size, the matrix $\bar{H}$ can be ill-conditioned, preventing a stable solution of the eigenvalue problem. In order to stabilize the algorithm, we add a small positive constant $\varepsilon$ to the diagonal matrix elements of $\bar{H}$ except for the first one, $\bar{H}_{ij} \to \bar{H}_{ij} + \varepsilon\,(1-\delta_{i0})\,\delta_{ij}$. This reduces the length of $\Delta\bar{{\bf p}}$ and it rotates it toward the steepest-descent direction.

Strong nonlinearities in the variational parameters potentially make $|\bar{\Psi}_T^\text{lin}({\bf p})\rangle$ significantly different from $|\bar{\Psi}_T({\bf p}^0+\Delta{\bf p})\rangle$. To alleviate this problem we employ the heuristic procedure of \cite{Contessi:2017}. For a given value of $\varepsilon$, equation~(\ref{eq:lin_sys}) is solved. If the linear variation of the trial state for ${\bf p}={\bf p}^0 +\Delta {{\bf p}}$ is small,  
\begin{align}
\frac{| \bar{\Psi}_T^\text{lin}({\bf p})|^2}{| \bar{\Psi}_T({\bf p}^0)|^2}=
1 + \sum_{i,j=1}^{N_p} \bar{S}_{ij}\,\Delta{p}^{i}\,\Delta{p}^{j} \le \delta ,
\label{eq:lin_var}
\end{align}
a short correlated run is performed in which the energy expectation value is estimated along the full variation of the trial state for a set of possible values of $\varepsilon$ (typically $50$ values are considered). The optimal $\varepsilon$ is the one corresponding to the lowest eigenvalue, provided that 
\begin{align}
\frac{|\bar{\Psi}_T({\bf p})|^2}{|\bar{\Psi}_T({\bf p}^0)|^2} \le \delta .
\label{eq:full_var}
\end{align}
Note that, in latter expression, at variance to equation~(\ref{eq:lin_var}), the full trial state instead of its linearized approximation appears in the numerator. This additional constraint suppresses the potential instabilities caused by the nonlinear dependence of the trial state on the variational parameters. When using the ``standard'' version of the LM, there were instances in which, despite the variation of the linear trial state being well below the threshold of equation~(\ref{eq:lin_var}), the full trial state fluctuated significantly more, preventing the convergence of the minimization algorithm. We found that choosing $\delta=0.2$ guarantees an ideal compromise between the convergence-rate and the stability of the  algorithm.  

When using the LM in conjunction with a density dependent interaction, one has to take into account that the parameters will now become themselves density dependent, as a consequence of the presence of the term $v^{\rm ddp}_{ij}$. This needs to be implemented in the calculation of the derivative of the local energy $E_L^i(X)$. To this aim, for each iteration of the LM, we first perform a preliminary Monte Carlo run, in which we compute the $2\,\mathcal N_p+1$ single-nucleon densities corresponding to the current set of variational parameters plus those with $p_i\to p_i + \delta_p$ and $p_i\to p_i - \delta_p$. We have found that this procedure reduces the benefit of using the non-hermitian estimator of $H_{ij}$ of equation~(\ref{eq:lm_est}). Indeed, explicitly symmetrizing $H_{ij}$ by means of $H_{ij}\to H_{ij}+H^*_{ji}$ does not significantly alter the convergence of the algorithm.

\section{Results}
\label{sec:res}
\subsection{ Comparison between HF and VMC}
\begin{figure}[t]
\centering
\includegraphics[width=0.5\textheight]{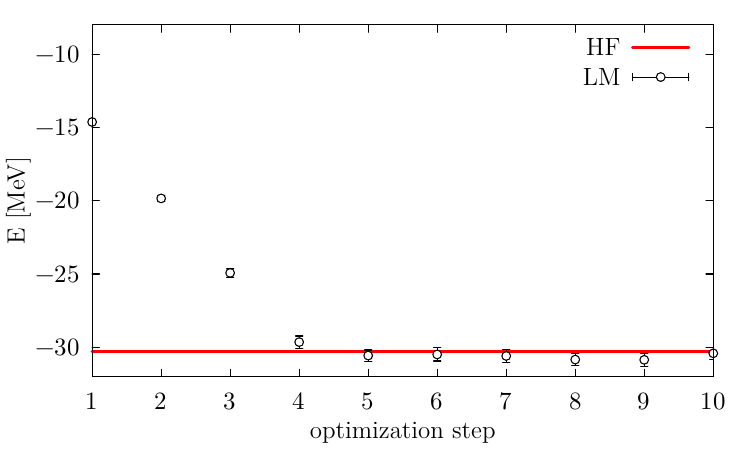}
\caption[]{Convergence pattern of the \isotope[4]{He} variational energy as a function of the number of optimization steps for the LM. As a comparison, the red line indicates the HF result.}
\label{fig:he4_lm}
\end{figure}

We first carried out a comparison between HF and VMC approaches by optimizing the single-particle orbitals of \isotope[4]{He} without introducing correlations 
in the VMC wave function. In both methods, we applied the same CM corrections, consisting in the subtraction of one- and two-body kinetic terms as in equation~(\ref{eq:ke_cm}). Hence, for this particular comparison, in the VMC calculation we refrain from subtracting the CM coordinates as in equation~(\ref{eq:r_i}). The convergence of the LM, shown in figure~\ref{fig:he4_lm}, is fast. Already after 10 optimization steps, the LM converges to the HF energy.

%In~\cref{tab:4he_res,tab:16o_res} 
In table~\ref{tab:4he-16o} we compare HF and LM results in \isotope[4]{He} and \isotope[16]{O} for the total energy $E$, the kinetic energy $T$, the expectation value of the different contributions of the potential of equation~(\ref{eq:gogny}), and the point-nucleon radius. The latter is defined as
\begin{align}
	\left\langle r_N^2\right\rangle=\frac{1}{{\cal N}}\big\langle\Psi\big|\sum_{i=1}^{A}\mathcal P_{N_i} |{\bf r}_i|^2\big|\Psi\big\rangle,
\end{align}
where ${\cal N}$ is the number of protons or neutrons,  
\begin{align}
	\mathcal P_{N_i}=\frac{1\pm\tau_{z_i}}{2},
	\label{eq:proj}
\end{align}
is the projector operator onto protons or neutrons, and ${\bf r}_i$ is the spatial rescaled coordinate defined in equation~(\ref{eq:r_i}).

\begin{table}[b]
\centering
\caption[]{Energy contributions (in MeV) and point-nucleon radii (in fm$^2$) in \isotope[4]{He} and \isotope[16]{O} obtained with HF and the LM. The experimental binding energy and point-proton radius are also reported.}
\begin{tabular}{l c ccc |c ccc}
                && \multicolumn{3}{c}{\isotope[4]{He}}  && \multicolumn{3}{c}{\isotope[16]{O}} \\
                && HF        & LM            & Exp      && HF        & LM          & Exp \\
\hline
$E$             && $-30.30$  & $-30.74(10)$  & $-28.30$ && $-129.1$  & $-128.8(5)$ & $-127.6$\\
$T$             && $39.45$   & $39.75(1)$    &          && $ 228.6$  & $227.9(1)$  & \\
$v_4+v_{\rm C}$ && $-132.91$ & $-133.87(13)$ &          && $-708.2$  & $-705.6(2)$ & \\
$v_{\rm ddp}$   && $ 63.16$  & $ 63.38(10)$  &          && $ 353.5$  & $348.9(5)$  & \\
\hline
$\langle r_{\rm pt}^2\rangle_p$ && $3.60$ & $3.60(1)$ & $2.14$~\cite{Sick:2008} && $7.10$ & $7.15(1)$ & $6.77$~\cite{Sick:1970} \\
$\langle r_{\rm pt}^2\rangle_n$ && $3.58$ & $3.57(1)$ &                         && $7.00$ & $7.05(1)$ & \\
\end{tabular}
\label{tab:4he-16o}
\end{table}   

The agreement between HF and LM results is good, for both energies and radii. The reason for the small discrepancies is mainly due to the fact that in our VMC calculations we use the regularized version of equation~(\ref{eq:ddp_reg}) rather than a pure contact density-dependent interaction. Note that, in \isotope[16]{O}, the deviation between the experimental energy ($-127.6\,\rm MeV$)  and the HF result  ($-129.1\,\rm MeV$) is $\approx2\,\rm MeV$. The HF value is consistent with the result of \cite{Arzhanov:2016} ($129.6\,\rm MeV$) once taking into account that the spin-orbit term of the Gogny interaction, not included in the present calculations, gives a contribution of about $0.7\,\rm MeV$ to the binding energy of \isotope[16]{O} (its contribution to the point-nucleon radius is negligible).

\begin{figure}[t]
\centering
\includegraphics[width=0.5\textheight]{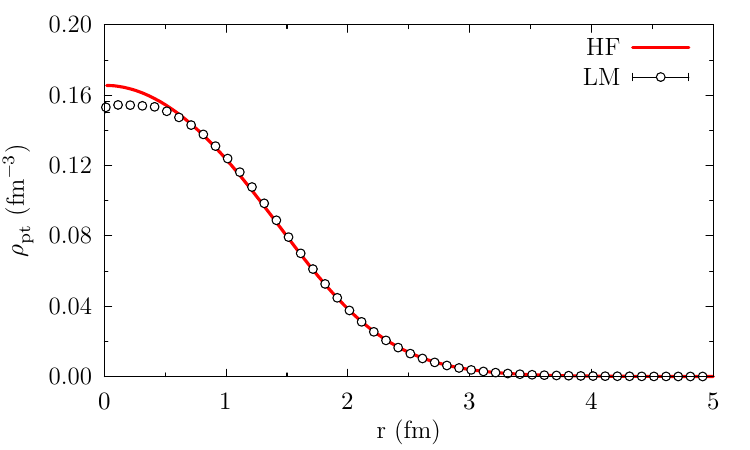}
\caption[]{Total point-nucleon density (protons plus neutrons) in \isotope[4]{He}.}
\label{fig:4he_rho}
\end{figure}

\begin{figure}[t]
\centering
\includegraphics[width=0.5\textheight]{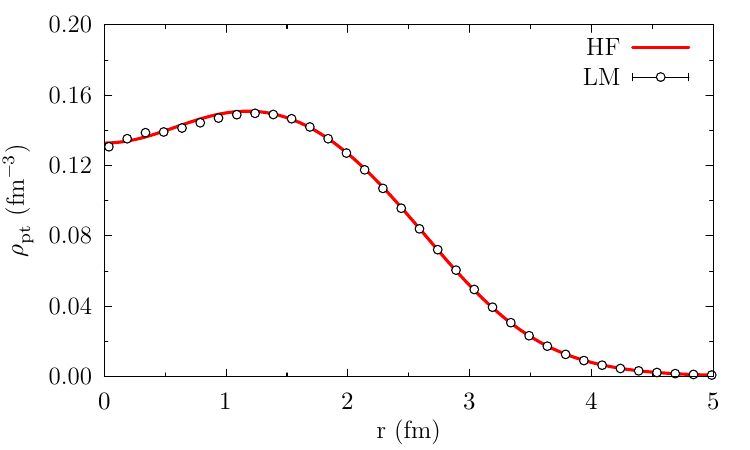}
\caption[]{Same of figure~\ref{fig:4he_rho} for \isotope[16]{O}.}
\label{fig:16o_rho}
\end{figure}

In figures~\ref{fig:4he_rho} and \ref{fig:16o_rho} we show the total HF and LM one-body point-nucleon density (protons plus neutrons) in \isotope[4]{He} and \isotope[16]{O}, respectively. Similarly to the point-nucleon radius, the one-body point-nucleon density is defined as
\begin{align}
\rho_{N}(r) =\frac{1}{4\pi r^2}\big\langle\Psi\big|\sum_{i=1}^{A}\mathcal P_{N_i}\delta(r-|{\bf r}_i|)\big|\Psi\big\rangle, 
\label{eq:rho_N}
\end{align}
where $\mathcal P_{N_i}$ is the projector operator of equation~(\ref{eq:proj}) and $\rho_N$ integrates to the number of nucleons. 
In QMC calculations the above expression is implemented by constructing the histogram of the nucleon radial coordinates $r_i$ multiplied by the expectation value of the projector operator, normalized to the number Monte Carlo samples and the volume element centered in $r_i$
\begin{align}
\rho_N(r)=\frac{1}{\mathcal M}\sum_{i=1}^{A}\frac{{\rm hist}(r_i)\big\langle\Psi\big|\mathcal P_{N_i}\big|\Psi\big\rangle}{\frac{4}{3}\pi\left[\left(r_i+\frac{dr}{2}\right)^3-\left(r_i-\frac{dr}{2}\right)^3\right]},
\label{eq:rho_hist}
\end{align}
where $dr$ is the size of the bin. Note that, since the nucleon coordinates are sampled from a probability distribution constructed using the trial wave function, and the latter is also used to calculate the expectation value of the projection operator, the density of equation~(\ref{eq:rho_N}) is always constructed for the actual wave function employed in the Monte Carlo run (correlated or uncorrelated). 

The agreement between the HF and LM densities is excellent, proving once again the accuracy of the LM in optimizing the radial components of the wave function.

\subsection{Center of mass effects}
To gauge CM contamination in HF calculations, analogously to what done in \cite{Rocco:2018}, we performed VMC calculations for \isotope[4]{He} and \isotope[16]{O} in the COMF, i.e. subtracting the CM coordinate as in equation~(\ref{eq:r_i}). Furthermore, we modify the density-dependent term of the Gogny potential, which is manifestly not Galilean invariant~\cite{Schmid:2002b}, by changing the definition of $v^{\rm ddp}_{ij}$ in equation~(\ref{eq:gogny_2}) to
\begin{align}
\rho^\alpha({\bf R}_{ij}) \to \rho^\alpha({\bf R}_{ij}-{\bf R}_{\rm CM}).
\end{align} 
In this way, Galilean invariance can be completely restored.

As a first step, we used the single-particle orbitals optimized for the comparative study between HF and LM, performing a PAV calculation. Results in the COMF are reported in the first column of tables~\ref{tab:4he} and \ref{tab:16o}, to be compared to those obtained in table~\ref{tab:4he-16o}. As one could have expected, the point-nucleon radii are much smaller when the CM motion is removed from the wave function. Correcting for the one- and two-body terms of equation~(\ref{eq:ke_cm}) takes care for most of the CM effects in the kinetic energy in the FRF. Note that the expectation values of these two terms exactly cancel in the COMF, as the CM kinetic energy vanishes in this reference frame. Although CM contamination mainly affects the expectation value of $v_{\rm ddp}$, the expectation values of $v_4+v_{\rm C}$ do also change from the FRF to the COMF. This might be somewhat surprising, as these terms only depend on the relative coordinate of the nucleon pair.

In order to clarify this point, let us consider the Hamiltonian~\cite{Giraud:2008}
\begin{equation}
H = H_{\rm CM} + H_{\rm int},
\label{eq:hsum}
\end{equation}
where the first term only depends on ${\bf R}_{\rm CM}$, while the second term is a function of the Jacobi coordinates. The latter, defined as 
\begin{align}
{\boldsymbol \xi}_1&={\bf r}_2-{\bf r}_1, \nonumber\\
{\boldsymbol \xi}_{2}&={\bf r}_3-\frac{{\bf r}_1 + {\bf r}_2}{2}, \nonumber\\
&\dots \nonumber \\
{\boldsymbol \xi}_{A-1}&={\bf r}_A-\frac{{\bf r}_1 + \dots + {\bf r}_{A-1}}{A-1} ,
\end{align}
are invariant under the CM subtraction of equation~(\ref{eq:r_i}). The ground-state wave function corresponding to the Hamiltonian of equation~(\ref{eq:hsum}) factorizes as 
\begin{equation}
\Psi_0({\bf r}_1,\dots,{\bf r}_A)=\Psi_{\rm CM}({\bf R}_{\rm CM})\Psi_{\rm int}({\boldsymbol \xi}_1,\dots,{\boldsymbol \xi}_{A-1}).
\label{eq:factor}
\end{equation}
Hence, the expectation value of any intrinsic operator $\mathcal O_{\rm int} ({\boldsymbol \xi}_{1},\dots,{\boldsymbol \xi}_{A-1})$ is independent of the CM coordinates
\begin{align}
\langle\mathcal O_{\rm int} \rangle =& \dfrac{\int d{\bf r}_{1} \dots d{\bf r}_{A}\Psi_0^\dagger({\bf r}_i)\,\mathcal O_{\rm int}\Psi_0({\bf r}_j)}{\int d{\bf r}_{1} \dots d{\bf r}_{A}|\Psi_0|^2} & i,j=1,\ldots,A\phantom{,} \nonumber\\[0.2cm]
 =& \dfrac{\int d{\boldsymbol \xi}_{1}\dots d{\boldsymbol \xi}_{A-1}\Psi^\dagger_{\rm int}(\boldsymbol\xi_l)\,\mathcal O_{\rm int}\Psi_{\rm int}(\boldsymbol\xi_m)}
{\int d{\boldsymbol \xi}_{1} \dots d{\boldsymbol \xi}_{A-1} |\Psi_{\rm int }|^2} & l,m=1,\ldots,A-1,
\label{eq:factorization}
\end{align}
as the factors $\int d{\bf R}_{\rm CM}|\Psi_{\rm CM}|^2$ from the numerator and the denominator simplify. However, since the density-dependent Hamiltonian of the Gogny interaction cannot be written as in equation~(\ref{eq:hsum}), the factorization of equation~(\ref{eq:factor}) does not apply and the expectation value of the intrinsic operator $v_4+v_{\rm C}$ might change when computed in the FRF or in the COMF.

As a second step, we performed a VAP calculation by optimizing the single-particle orbitals of \isotope[4]{He} and \isotope[16]{O} with the LM in the COMF. To compensate for the increased density-dependent term, the optimization procedure broadens the single-particle orbitals. As a consequence, in the results shown in the second column of tables~\ref{tab:4he} and \ref{tab:16o}, the point-nucleon radii are larger than those obtained in the FRF, see table table~\ref{tab:4he-16o}. Consistently, the expectation values of the kinetic energy and of the separate potential terms are also smaller (in absolute values). These effects are, as expected, more prominent in \isotope[4]{He}, although both nuclei are appreciably under-bound. The effect of CM projection on the binding energies of $^4$He and \isotope[16]{O} can be obtained comparing the second column of tables~\ref{tab:4he} and \ref{tab:16o} with the HF results reported in table~\ref{tab:4he-16o} and in figure~\ref{fig:he4_lm}. It accounts for $10.2\,\rm MeV$ in $^4$He and $9.6\,\rm MeV$ in \isotope[16]{O}. These values are in good agreement with those obtained by Schmid (see figure~7 in \cite{Schmid:2002b}).

\begin{table}[t]
\centering
\caption[]{Energy contributions (in MeV) and point-nucleon radii (in fm$^2$) in \isotope[4]{He}. First column: PAV calculation (VMC calculations in the COMF using the same single-particle orbitals of table~\ref{tab:4he-16o}); Second column: VAP calculation; Third column: VAP$^*$ calculation (renormalized $\tau_0^*$). The experimental binding energy and point-proton radius are also reported in the last column.} 
\begin{tabular}{lccccc}
                 & HF         & PAV          & VAP         & VAP$^*$      & Exp \\
\hline
$E$              & $-30.30$ & $-15.93(9)$  & $-20.12(8)$ & $-28.31(8)$  & $-28.30$ \\
$T$              & $39.45$  & $39.42(1)$   & $28.49(1)$  & $35.84(1)$   & \\
$v_4+v_{\rm C}$  & $-132.91$& $-125.57(1)$ & $-97.66(1)$ & $-126.24(1)$ & \\
$v_{\rm ddp}$    & $63.16$  & $70.22(9)$   & $ 49.06(8)$ & $ 62.09(8)$  & \\
\hline
$\langle r_{\rm pt}^2\rangle_p$ &$3.60$ & $2.70(1)$ & $4.11(1)$ & $3.20(1)$ & $2.14$~\cite{Sick:2008} \\ 
$\langle r_{\rm pt}^2\rangle_n$ &$3.58$ & $2.69(1)$ & $4.09(1)$ & $3.19(1)$ & \\
\end{tabular}
\label{tab:4he}
\end{table}

\begin{table}[t]
\centering
\caption[]{Same as table~\ref{tab:4he} for \isotope[16]{O}.} 
\begin{tabular}{lccccc}
                 &HF        & PAV          & VAP         & VAP$^*$    & Exp \\
\hline                     
$E$             & $-129.1$ & $-118.6(3)$ & $-119.5(3)$ & $-127.7(2)$ & $-127.6$ \\
$T$             & $228.6$  & $228.1(1)$  & $221.6(1)$  & $232.9(1)$  & \\
$v_4+v_{\rm C}$ & $-708.2$ & $-703.8(1)$ & $-684.3(1)$ & $-731.5(1)$ & \\
$v_{\rm ddp}$   & $353.5$  & $357.1(4)$  & $ 343.2(3)$ & $ 370.9(2)$ & \\
\hline
$\langle r_{\rm pt}^2\rangle_p$ & $7.10$ & $6.83(1)$ & $7.12(1)$ & $6.78(1)$ & $6.77$~\cite{Sick:1970} \\
$\langle r_{\rm pt}^2\rangle_n$ & $7.00$ & $6.78(1)$ & $7.07(1)$ & $6.70(1)$ & \\
\end{tabular}
\label{tab:16o}
\end{table}

The set of parameters of the Gogny force we have employed so far are meaningful only if the system is described in the FRF. However, it is possible, in principle, to refit such parameters in order to reproduce observables in the COMF. We leave this task to a future work. Here, we limit ourselves to reducing the strength of the density-dependent term of the Gogny interaction by renormalizing the coefficient $\tau_0$ of equation~(\ref{eq:gogny_2}) ($1390.6$ MeV) so as to reproduce the experimental binding energies in the case of \isotope[4]{He} and \isotope[16]{O}. Since CM effects in $v^{\rm ddp}_{ij}$ are larger in lighter nuclei, the quenching is stronger in \isotope[4]{He} ($\tau_0^*=1192.4\,\rm MeV$) than in \isotope[16]{O} ($\tau_0^*=1357.5\,\rm MeV$), consistently with the findings of \cite{Schmid:2002b}. The binding energies and radii of \isotope[4]{He} and \isotope[16]{O} obtained employing the renormalized value of $\tau_0^*$, denoted as VAP$^*$, are listed in the third column of tables~\ref{tab:4he} and \ref{tab:16o}. 

\begin{figure}[t]
\centering
\includegraphics[width=0.5\textheight]{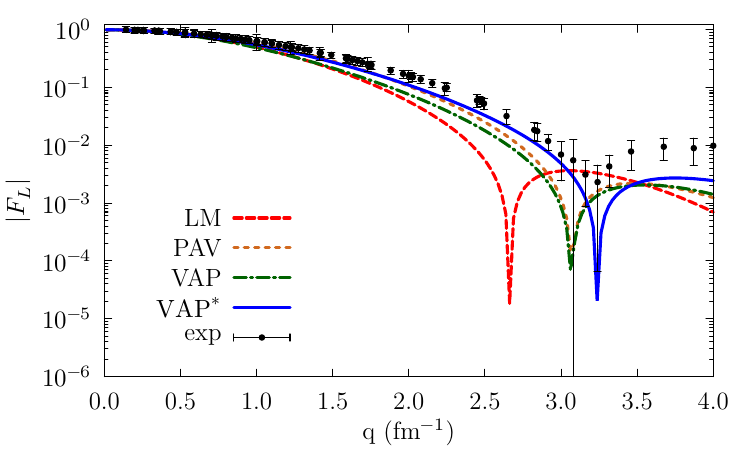}
\caption[]{Longitudinal form factor in~\isotope[4]{He}. Statistical Monte Carlo uncertainties are smaller than the thickness of the lines. Experimental data are from I. Sick~\cite{Sick:2005}, based on \cite{Frosch:1967,Erich:1968,McCarthy:1977,Arnold:1978,Ottermann:1985}.}
\label{fig:4he_ff}
\end{figure}

\begin{figure}[t]
\centering
\includegraphics[width=0.5\textheight]{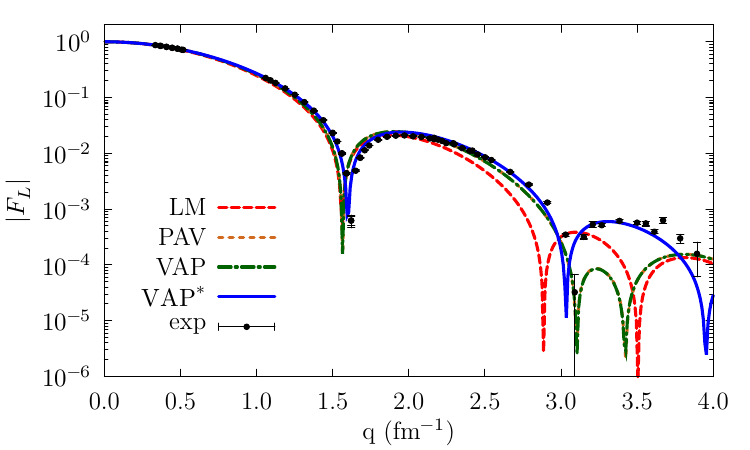}
\caption[]{Same of figure~\ref{fig:4he_ff} for \isotope[16]{O}. Experimental data are from I. Sick~\cite{Sick:2005}, based on \cite{Sick:1970,Schuetz:1975,Sick:1975}}
\label{fig:16o_ff}
\end{figure}

The quenching of the repulsive density-dependent term of equation~(\ref{eq:gogny}) implies a reduction of the point-nucleon radii, which is reflected in the longitudinal elastic form factors (charge form factors), as shown in figures~\ref{fig:4he_ff} and \ref{fig:16o_ff}. The charge form factor is expressed as the ground-state expectation value of the one-body charge operator~\cite{Mcvoy:1962}
\begin{align}
	F_L(q)=\frac{1}{Z}\frac{G_E^p(Q_{\rm el}^2)\,\tilde{\rho}_p(q)+G_E^n(Q_{\rm el}^2)\,\tilde{\rho}_n(q)}{\sqrt{1+Q_{\rm el}^2/(4 m_N^2)}} ,
	\label{eq:ff}
\end{align}
where $\tilde{\rho}_{N}(q)$ is the Fourier transform of the one-body point-nucleon density defined in equation~(\ref{eq:rho_N}), and $Q^2_{\rm el}={\bf q}^2-\omega_{\rm el}^2$ is the four-momentum squared, with $\omega_{\rm el}=\sqrt{q^2+m_A^2}-m_A$ the energy transfer corresponding to the elastic peak, $m_A$ being the mass of the target nucleus. $G_E^N(Q^2)$ are the nucleon electric form factors, for which we adopt Kelly's parametrization~\cite{Kelly:2004}. The above expression is derived ignoring small spin-orbit contributions in the one-body current.

In \isotope[4]{He}, subtracting the CM contributions and using the quenched $\tau_0^*$ coefficient significantly improves the agreement with experimental data. As discussed in \cite{Carlson:2015,Marcucci:2016,Lynn:2017}, meson-exchange currents are needed to shift the peaks of the longitudinal elastic form factor to lower values of the momentum transfer and achieve agreement with experiment. On the other hand, the fact that experimental data at small $q$ are underestimated, and consequently the proton radius is overpredicted, suggests that the simple quenching of $\tau_0$ does not suffice to adequately describe \isotope[4]{He}.  

CM spuriosities have less effect on the longitudinal form factor of \isotope[16]{O}. Even in this case, however, working in the COMF significantly improves the agreement with experiment in the whole momentum-transfer region. In particular, our VAP$^*$ calculations reproduce very well both the first and the second diffraction peaks, despite the Gogny D1S interaction is not expected to be reliable in this kinematical regime.

Finally we remark  that the trends observed in figures~\ref{fig:4he_ff} and \ref{fig:16o_ff}, and in particular the differences between the momentum dependence of LM and VAP results, are qualitatively similar to those calculated in \cite{Rodriguez:2004a}, although quantitative statements are not possible since a simplified interaction was adopted in \cite{Rodriguez:2004a}.

\begin{figure}[h]
\centering
\includegraphics[width=0.5\textheight]{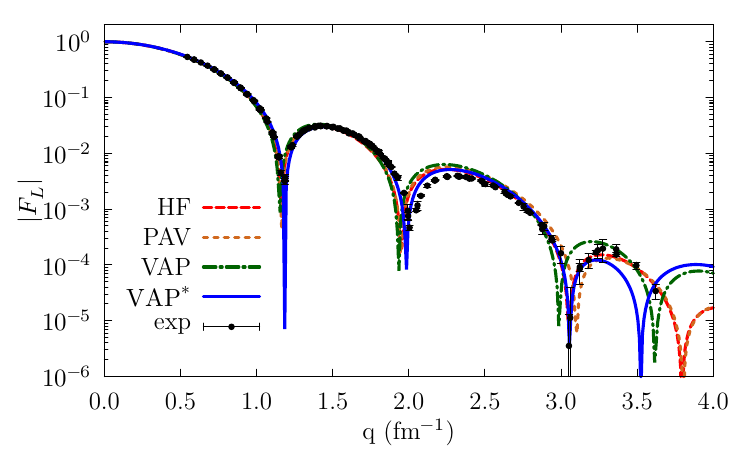}
\caption[]{Same of figure~\ref{fig:4he_ff} for \isotope[40]{Ca}. Experimental data are from I. Sick~\cite{Sick:2005}, based on \cite{Sinha:1973,Sick:1975,Sick:1979}}
\label{fig:40ca_ff}
\end{figure}

We complete our analyses by studying a heavier nucleus, namely \isotope[40]{Ca}. In this case, in order to save computing time, we did not perform LM calculations and we used instead the single-particle orbitals obtained from HF as a starting point for obtaining the PAV, VAP, and VAP$^*$ results. As shown in table~\ref{tab:40ca}, the main differences between the HF and PAV binding energies originate from the density-dependent contribution, whose increase can be directly related to the smaller PAV radii. On the other hand, the HF and PAV expectation values of $v_4+v_C$ are very similar, reflecting the fact that the factorization of equation~(\ref{eq:factorization}) is approximately fullfilled for $^{40}$Ca.  Consistently with \isotope[4]{He} and \isotope[16]{O}, in order to reduce the contribution of $v_{ddp}$, the LM in the COMF broadens the single-particle orbitals, so that the VAP radii are larger than the PAV ones. In order for VAP$^*$ results to match the experimental binding energy, a milder quenching with respect to the original value, $\tau_0^*=1390.0\,\rm MeV$, is needed in the case of \isotope[40]{Ca} ($\tau_0^*=1378.1\,\rm MeV$) in comparison to \isotope[4]{He} ($\tau_0^*=1192.4\,\rm MeV$) and \isotope[16]{O} $(\tau_0^*=1357.5\,\rm MeV)$. As expected, this procedure yields to smaller point radii, which are in better agreement with the experimental values. A better reproduction of experimental data for the longitudinal form factor, displayed in figure~\ref{fig:40ca_ff}, is also observed, with the first, second, and third diffraction peaks correctly reproduced.

\begin{table}[]
\centering
\caption[]{Same as table~\ref{tab:4he} for \isotope[40]{Ca}.} 
\begin{tabular}{lccccc}
                 & HF        & PAV   & VAP   & VAP$^*$ & Exp \\
\hline                     
$E$              & $-343.3$  & $-334.7(9)$  & $-334.8(8)$  & $-342.3(9)$ & $-342.1$ \\
$T$              & $ 642.6$  & $642.6(1)$   & $636.0(1)$   & $643.9(1)$ & \\
$v_4+v_{\rm C}$  & $-2038.6$ & $-2038.7(1)$ & $-2005.6(1)$ & $-2038.9(1)$ & \\
$v_{\rm ddp}$    & $1052.7$  & $1061.4(9)$  & $1034.8(9)$ & $1052.7$ & \\
\hline
$\langle r_{\rm pt}^2\rangle_p$ & $11.68$ & $11.52(1)$ & $11.66(1)$ & $11.51(1)$ & $11.41$~\cite{Wohlfahrt:1981} \\
$\langle r_{\rm pt}^2\rangle_n$ & $11.37$ & $11.22(1)$ & $11.36(1)$ & $11.21(1)$ & \\
\end{tabular}
\label{tab:40ca}
\end{table}

\section{Conclusions}
\label{sec:con}
We have carefully analyzed the effects of violating Galilean invariance in mean-field calculations based on density-dependent nuclear effective interactions. To this aim, we have devised a novel variational Monte Carlo approach in which the optimal single-particle orbitals are found using a generalization of the Linear Method, suitable to treat interactions that explicitly depend on the density of the system. Although we have only considered the widely used D1S parametrization of the Gogny potential, our analysis remains valid for other density-dependent interactions.

A rigorous way of singling-out CM effects when solving the nuclear Schr\"odinger equation consists in employing the Jacobi coordinates, as done in standard few-body techniques, such as the hyperspherical harmonics method~\cite{Barnea:1999be,Viviani:2005gu}. Because of the factorial scaling of the required antisymmetrization, using Jacobi coordinates becomes impractical as the number of nucleons increases, and it is currently limited to few-body problems with $A\lesssim 6$. For larger systems, CM contributions are removed {\it a posteriori}, exploiting the factorization of the nuclear wave function in an intrinsic and in a CM state~\cite{Hagen:2009pq}. However, as we have shown in this work, the density dependence of the effective interaction brings about potential violations to this factorization, making the latter approach unfeasible. 
 
QMC methods are ideally suited for an exact treatment of the CM problem in nuclear many-body calculations, even when density-dependent interactions are used. In fact, within QMC one can simultaneously access all the $3A$ single-particle coordinates of the nucleons, and the CM coordinate can be easily subtracted, as in equation~(\ref{eq:r_i}). As a consequence, the density-dependent term of the nuclear interaction is automatically modified in a way that does no longer violate Galilean invariance. The excellent agreement between this method and techniques based on the Jacobi coordinates~\cite{Kamada:2001tv} corroborates the validity of the QMC treatment of the CM motion.
 
At first, we have benchmarked our QMC results against accurate HF calculations for two closed-shell nuclei, namely \isotope[4]{He} and \isotope[16]{O}. For the sake of this comparison, in both methods the violation of translational invariance is only partially accounted for by correcting the kinetic energy contribution. We have found that QMC and HF results are consistent, hence demonstrating the reliability of our improved version of the LM.   

Then, imposing that the HF orbitals are not modified, we found that QMC calculations in the COMF (PAV) yield underbound nuclei, but the radii are in closer agreement with experiments. Optimizing the uncorrelated wave functions in the COMF (VAP) leads instead to underbinding and to an overestimate of the radii, compared to the PAV calculations. In order to recover the experimental results, we have renormalized the repulsive term in the Gogny force (VAP$^*$ calculations). For this subsequent analysis we have also considered the \isotope[40]{Ca} nucleus. The required quenching decreases with the mass of the nucleus, indicating that the mean-field description is less and less dependent on CM effects for larger nuclei, as expected. The same analysis has been extended to the charge form factors, leading to consistent conclusions. As for the latter quantity, our VAP$^*$ results for \isotope[16]{O} and \isotope[40]{Ca} turn out to be in excellent agreement with experimental data.  

This work opens the way of a more systematic account of Galilean invariance, when density dependent interactions are used. The quenching applied to the repulsive component of the Gogny interaction to reproduce the binding energy of \isotope[16]{O} is likely to underbind heavier nuclear systems, where CM corrections are less relevant. This calls for a global refit of the parameters of the Gogny interaction, including both \isotope[16]{O} and larger nuclei, such as \isotope[40]{Ca} and \isotope[48]{Ca}.

\ack{We thank G. Co', J. Margueron, A. Rios, X. Roca-Maza, and A. Roggero for valuable discussions. F.B and E.V. acknowledge funding from the European Union Horizon 2020 research and innovation program under Grant Agreement No. 654002. F.B. acknowledges funding from the Spanish Ministerio de Econom\'\i a under Grant Agreement No. FIS2017-88410-P. The work of D.L. was supported by the U.S. Department of Energy, Office of Science, Office of Nuclear Physics, under Contract No. DE-SC0013617, and by the NUCLEI SciDAC program. The work of A.L. was supported by the U.S. Department of Energy, Office of Science, Office of Nuclear Physics, under Contract No. DE-AC02-06CH11357. Numerical calculations have been made possible through a CINECA-INFN agreement, providing access to resources on MARCONI at CINECA. We gratefully acknowledge the computing resources provided on Bebop, a high-performance computing cluster operated by the Laboratory Computing Resource Center at Argonne National Laboratory. This research also used resources of the Argonne Leadership Computing Facility, which is a DOE Office of Science User Facility supported under Contract No. DE-AC02-06CH11357.}  

%\bibliographystyle{iopart-num}
%\bibliography{biblio}

\begin{thebibliography}{10}
\expandafter\ifx\csname url\endcsname\relax
  \def\url#1{{\tt #1}}\fi
\expandafter\ifx\csname urlprefix\endcsname\relax\def\urlprefix{URL }\fi
\providecommand{\eprint}[2][]{\url{#2}}

\bibitem{Witala:2001}
Wita\l{}a H, Gl\"ockle W, Golak J, Nogga A, Kamada H, Skibi\'{n}ski R and
  Kuro\'s-\.{Z}o\l{}nierczuk J 2001 {\em Phys. Rev. C\/} {\bf 63}(2) 024007

\bibitem{Lazauskas:2004}
Lazauskas R and Carbonell J 2004 {\em Phys. Rev. C\/} {\bf 70}(4) 044002

\bibitem{Lazauskas:2009}
Lazauskas R 2009 {\em Phys. Rev. C\/} {\bf 79}(5) 054007

\bibitem{Deltuva:2007}
Deltuva A and Fonseca A~C 2007 {\em Phys. Rev. C\/} {\bf 75}(1) 014005

\bibitem{Kievsky:2008}
Kievsky A, Rosati S, Viviani M, Marcucci L and Girlanda L 2008 {\em J. Phys. G
  Nucl. Part. Phys.\/} {\bf 35} 063101

\bibitem{Marcucci:2009}
Marcucci L~E, Kievsky A, Girlanda L, Rosati S and Viviani M 2009 {\em Phys.
  Rev. C\/} {\bf 80}(3) 034003

\bibitem{Navratil:2000}
Navr\'atil P, Vary J~P and Barrett B~R 2000 {\em Phys. Rev. Lett.\/} {\bf
  84}(25) 5728--5731

\bibitem{Barrett:2013}
Barrett B, Navr\'atil P and Vary J 2013 {\em Prog. Part. Nucl. Phys.\/} {\bf
  69} 131--181 ISSN 0146-6410

\bibitem{Baroni:2013}
Baroni S, Navr\'atil P and Quaglioni S 2013 {\em Phys. Rev. Lett.\/} {\bf
  110}(2) 022505

\bibitem{Navratil:2016}
Navr\'atil P, Quaglioni S, Hupin G, Romero-Redondo C and Calci A 2016 {\em
  Phys. Scr.\/} {\bf 91} 053002

\bibitem{Ring:1980}
Ring P and Schuck P 1980 {\em {The nuclear many-body problem}\/} (Berlin,
  Heidelberg: Springer Berlin Heidelberg)

\bibitem{Bender:2003}
Bender M, Heenen P~H and Reinhard P~G 2003 {\em Rev. Mod. Phys.\/} {\bf 75}(1)
  121--180

\bibitem{Schmid:2001}
{KW Schmid} 2001 {\em Eur. Phys. J. A\/} {\bf 12} 29--40

\bibitem{Schmid:2002a}
{KW Schmid} 2002 {\em Eur. Phys. J. A\/} {\bf 13} 319--338

\bibitem{Schmid:2002b}
{KW Schmid} 2002 {\em Eur. Phys. J. A\/} {\bf 14} 413--438

\bibitem{Schmid:2003}
Schmid K~W 2003 {\em Eur. Phys. J. A\/} {\bf 16} 475--487 ISSN 1434-601X

\bibitem{Rodriguez:2004a}
Rodriguez-Guzman R and Schmid K~W 2004 {\em Eur. Phys. J. A\/} {\bf 19} 45--59

\bibitem{Rodriguez:2004b}
Rodriguez-Guzman R and Schmid K~W 2004 {\em Eur. Phys. J. A\/} {\bf 19} 61--75

\bibitem{Decharge:1979}
Decharg\'e J and Gogny D 1980 {\em Phys. Rev. C\/} {\bf 21} 1568--1593

\bibitem{Robledo:2018cdj}
Robledo L~M, Rodr\'iguez T~R and Rodr\'iguez-Guzm\'an R~R 2019 {\em J. Phys. G:
  Nucl. Part. Phys.\/} {\bf 46} 013001

\bibitem{Carlson:2015}
Carlson J, Gandolfi S, Pederiva F, Pieper S~C, Schiavilla R, Schmidt K~E and
  Wiringa R~B 2015 {\em Rev. Mod. Phys.\/} {\bf 87} 1067--1118

\bibitem{Toulouse:2007}
{Toulouse} J and {Umrigar} C~J 2007 {\em J. Chem. Phys.\/} {\bf 126} 084102

\bibitem{Contessi:2017}
Contessi L, Lovato A, Pederiva F, Roggero A, Kirscher J and van Kolck U 2017
  {\em Phys. Lett. B\/} {\bf 772} 839--848 ISSN 0370-2693

\bibitem{Berger:1991}
Berger J~F, Girod M and Gogny D 1991 {\em Comp. Phys. Comm.\/} {\bf 63}
  365--374

\bibitem{Wiringa:1995}
Wiringa R~B, Stoks V~G~J and Schiavilla R 1995 {\em Phys. Rev. C\/} {\bf 51}(1)
  38--51

\bibitem{Goutte:2005da}
Goutte H, Berger J~F, Casoli P and Gogny D 2005 {\em Phys. Rev. C\/} {\bf
  71}(2) 024316

\bibitem{RocaMaza:2011pm}
Roca-Maza X, Centelles M, Vi\~nas X and Warda M 2011 {\em Phys. Rev. Lett.\/}
  {\bf 106}(25) 252501

\bibitem{Gaffney:2013}
Gaffney L~P, Butler P~A, Scheck M, Hayes A~B, Wenander F, Albers M, Bastin B,
  Bauer C, Blazhev A, B{\"o}nig S, Bree N, Cederk{\"a}ll J, Chupp T, Cline D,
  Cocolios T~E, Davinson T, De~Witte H, Diriken J, Grahn T, Herzan A, Huyse M,
  Jenkins D~G, Joss D~T, Kesteloot N, Konki J, Kowalczyk M, Kr{\"o}ll T, Kwan
  E, Lutter R, Moschner K, Napiorkowski P, Pakarinen J, Pfeiffer M, Radeck D,
  Reiter P, Reynders K, Rigby S~V, Robledo L~M, Rudigier M, Sambi S, Seidlitz
  M, Siebeck B, Stora T, Thoele P, Van~Duppen P, Vermeulen M~J, von Schmid M,
  Voulot D, Warr N, Wimmer K, Wrzosek-Lipska K, Wu C~Y and Zielinska M 2013
  {\em Nature\/} {\bf 497} 199

\bibitem{Sellahewa:2014}
Sellahewa R and Rios A 2014 {\em Phys. Rev. C\/} {\bf 90}(5) 054327

\bibitem{Piarulli:2014}
Piarulli M, Girlanda L, Schiavilla R, P\'erez R~N, Amaro J~E and Arriola E~R
  2015 {\em Phys. Rev. C\/} {\bf 91}(2) 024003

\bibitem{Piarulli:2017}
Piarulli M, Baroni A, Girlanda L, Kievsky A, Lovato A, Lusk E, Marcucci L~E,
  Pieper S~C, Schiavilla R, Viviani M and Wiringa R~B 2018 {\em Phys. Rev.
  Lett.\/} {\bf 120}(5) 052503

\bibitem{Lovato:2011ij}
Lovato A, Benhar O, Fantoni S and Schmidt K~E 2012 {\em Phys. Rev. C\/} {\bf
  85}(2) 024003

\bibitem{Huth:2017}
Huth L, Tews I, Lynn J~E and Schwenk A 2017 {\em Phys. Rev. C\/} {\bf 96}(5)
  054003

\bibitem{Bohr:1969}
Bohr A and Mottelson B 1965 {\em {Nuclear structure}\/} (New York: Benjamin)

\bibitem{Briganti:1998}
Briganti S 1998 {Study of the structure of exotic nuclei with
  Hartree-Fock-Bogoliubov theory} {Master Thesis, University of Milano,
  unpublished}

\bibitem{Engel:2007}
Engel J 2007 {\em Phys. Rev. C\/} {\bf 75}(1) 014306

\bibitem{Messud:2011}
Messud J 2011 {\em Phys. Rev. A\/} {\bf 84}(5) 052113

\bibitem{Messud:2013}
Messud J 2013 {\em Phys. Rev. C\/} {\bf 87}(2) 024302

\bibitem{Lesinski:2014}
Lesinski T 2014 {\em Phys. Rev. C\/} {\bf 89}(4) 044305

\bibitem{Giraud:2008}
Giraud B~G 2008 {\em Phys. Rev. C\/} {\bf 77}(1) 014311

\bibitem{Metropolis:1953}
Metropolis N, Rosenbluth A~W, Rosenbluth M~N, Teller A~H and Teller E 1953 {\em
  J. Chem. Phys.\/} {\bf 21} 1087--1092

\bibitem{Ceperley:1995}
Ceperley D~M 1995 {\em Rev. Mod. Phys.\/} {\bf 67}(2) 279--355

\bibitem{Schmidt:1981}
Schmidt K~E, Lee M~A, Kalos M~H and Chester G~V 1981 {\em Phys. Rev. Lett.\/}
  {\bf 47}(11) 807--810

\bibitem{Umrigar:2005}
Umrigar C~J and Filippi C 2005 {\em Phys. Rev. Lett.\/} {\bf 94}(15) 150201

\bibitem{Motta:2015}
Motta M, Bertaina G, Galli D and Vitali E 2015 {\em Comp. Phys. Comm.\/} {\bf
  190} 62--71 ISSN 0010-4655

\bibitem{Nightingale:2001}
Nightingale M~P and Melik-Alaverdian V 2001 {\em Phys. Rev. Lett.\/} {\bf
  87}(4) 043401

\bibitem{Sick:2008}
Sick I 2008 {\em {Precise radii of light nuclei from electron scattering}\/}
  (Berlin: Springer) pp 57--77 ISBN 978-3-540-75479-4

\bibitem{Sick:1970}
Sick I and McCarthy J~S 1970 {\em Nucl. Phys.\/} {\bf A150} 631--654

\bibitem{Arzhanov:2016}
Arzhanov A, Rodr\'{\i}guez T~R and Mart\'{\i}nez-Pinedo G 2016 {\em Phys. Rev.
  C\/} {\bf 94}(5) 054319

\bibitem{Rocco:2018}
Rocco N and Barbieri C 2018 {\em Phys. Rev. C\/} {\bf 98}(2) 025501

\bibitem{Mcvoy:1962}
McVoy K~W and Van~Hove L 1962 {\em Phys. Rev.\/} {\bf 125}(3) 1034--1043

\bibitem{Kelly:2004}
Kelly J~J 2004 {\em Phys. Rev. C\/} {\bf 70}(6) 068202

\bibitem{Sick:2005}
Sick I 2005 {unpublished}

\bibitem{Frosch:1967}
Frosch R~F, McCarthy J~S, Rand R~E and Yearian M~R 1967 {\em Phys. Rev.\/} {\bf
  160}(4) 874--879

\bibitem{Erich:1968}
Erich U, Frank H, Haas D and Prange H 1968 {\em Z. Phys.\/} {\bf 209} 208--218
  ISSN 0939-7922

\bibitem{McCarthy:1977}
McCarthy J~S, Sick I and Whitney R~R 1977 {\em Phys. Rev. C\/} {\bf 15}(4)
  1396--1414

\bibitem{Arnold:1978}
Arnold R~G, Chertok B~T, Rock S, Sch\"utz W~P, Szalata Z~M, Day D, McCarthy
  J~S, Martin F, Mecking B~A, Sick I and Tamas G 1978 {\em Phys. Rev. Lett.\/}
  {\bf 40}(22) 1429--1432

\bibitem{Ottermann:1985}
Ottermann C~R, K\"obschall G, Maurer K, R\"ohrich K, Schmitt C and Walther V~H
  1985 {\em Nucl. Phys. A\/} {\bf 436} 688--698 ISSN 0375-9474

\bibitem{Schuetz:1975}
Sch{\"u}tz W 1975 {\em Z. Phys. A\/} {\bf 273} 69--75 ISSN 0939-7922

\bibitem{Sick:1975}
Sick I 1975 {unpublished}

\bibitem{Marcucci:2016}
Marcucci L~E, Gross F, Pena M~T, Piarulli M, Schiavilla R, Sick I, Stadler A,
  Van~Orden J~W and Viviani M 2016 {\em J. Phys. G Nucl. Part. Phys.\/} {\bf
  43} 023002

\bibitem{Lynn:2017}
Lynn J~E, Tews I, Carlson J, Gandolfi S, Gezerlis A, Schmidt K~E and Schwenk A
  2017 {\em Phys. Rev. C\/} {\bf 96}(5) 054007

\bibitem{Wohlfahrt:1981}
Wohlfahrt H~D, Shera E~B, Hoehn M~V, Yamazaki Y and Steffen R~M 1981 {\em Phys.
  Rev. C\/} {\bf 23}(1) 533--548

\bibitem{Sinha:1973}
Sinha B~B~P, Peterson G~A, Whitney R~R, Sick I and McCarthy J~S 1973 {\em Phys.
  Rev. C\/} {\bf 7}(5) 1930--1938

\bibitem{Sick:1979}
Sick I, Bellicard J, Cavedon J, Frois B, Huet M, Leconte P, Ho P and Platchkov
  S 1979 {\em Phys. Lett. B\/} {\bf 88} 245--248 ISSN 0370-2693

\bibitem{Barnea:1999be}
Barnea N, Leidemann W and Orlandini G 2000 {\em Phys. Rev. C\/} {\bf 61} 054001

\bibitem{Viviani:2005gu}
Viviani M, Marcucci L~E, Rosati S, Kievsky A and Girlanda L 2006 {\em Few Body
  Syst.\/} {\bf 39} 159--176

\bibitem{Hagen:2009pq}
Hagen G, Papenbrock T and Dean D~J 2009 {\em Phys. Rev. Lett.\/} {\bf 103}
  062503

\bibitem{Kamada:2001tv}
Kamada H, Nogga A, Gl\"ockle W, Hiyama E, Kamimura M, Varga K, Suzuki Y,
  Viviani M, Kievsky A, Rosati S, Carlson J, Pieper S~C, Wiringa R~B,
  Navr\'atil P, Barrett B~R, Barnea N, Leidemann W and Orlandini G 2001 {\em
  Phys. Rev. C\/} {\bf 64}(4) 044001

\end{thebibliography}
\providecommand{\newblock}{}

\end{document}